# Dynamic Rupture, Fault Opening, and Near Fault Particle Motions along an Interface between Dissimilar Materials

Baoping Shi[1], Yanheng Li[1], Jian Zhang[1] and James N. Brune[2]

[1] School of Earth Sciences, Graduate School of Chinese Academy of Sciences, Beijing Yuquan Road #19, Beijing 100049, China

[2] Seismological Laboratory, MD 174, University of Nevada, Reno, NV 89557-0141, USA

June 22, 2005

## Abstract

Dynamic rupture propagation along an interface between two different elastic solids under shear dominated loading is studied numerically by a 2-D lattice particle model (LPM). The configuration of the lattice particle model consists of two solid blocks of different elastic properties connected along a planar interface. Each block is characterized as an isotropic elastic material and the interface strength is described as a composite elastic modulus of a mismatch function of the elastic properties of two dissimilar materials. The particle interaction between the two blocks with pair inter-particle potential also takes account of normal stress variations.

Numerical simulations illustrate that, when an initiated rupture direction is the same as the slip direction of compliant material (softer), the dynamic rupture propagates with a self-sustaining slip pulse along the fault at the speed close to the slower *Rayleigh* wave speed, accompanied by a temporary and localized interface separation (or fault opening). The interface separation at a point on the fault is indicated by the fault normal displacement discontinuity between two blocks, while local dynamic shear stress drops to zero instantaneously. The normal particle motions at the two sides of the fault have the same direction, towards the softer material, and the particle velocity in the fault normal direction are much larger than that in the fault parallel direction. The observed particle motions are consistent with the foam rubber experiments and are very similar to the results predicted by the *Weertman's* dislocation theory, and the *Schallamach* waves as the material contrast exceeds 40%. Moreover, corresponding to the fault trace, the near fault particle motions are strong asymmetrical between soft and hard blocks, and the particle velocity in the softer material is larger than that in the harder material. In addition, the synthetic seismograms also revealed that the large particle motions both in the fault normal and parallel directions are contributed from the surface wave energy excited by the source rupture process. The radiated seismic energy comes from the particle slip/opening motions and healing process (stopping phase). Therefore, the shear stress variation on the fault behaves as a partial stress drop during the rupture process. The major frictional energy is associated with the work done in pulling the contact points apart as the rupture wave propagates.

## Introduction:

In spite of three more decades of study and speculation, the heat flow paradox, which is the lack of any indication of frictional heat generation along the San Andreas faults (Brune *et al*., 1969; Lachenbruch and Sass, 1980), remains unsolved. At present the most frequently discussed mechanism for explaining the heat flow paradox is the reduction of effective normal stress by near-lithostatic pore pressures (Rice, 1992), or by pods of high pore pressure generated and locked in by permeability valves (Byerlee, 1990, 1992). A major problem with the high pore pressure explanation is the question of how the pore pressure can be maintained at such high levels for hundred, up to tens of thousands of years.

New concepts for the process of stick-slip have the potential of not only explaining the heat flow paradox, but also explaining other puzzling features of earthquake slip, some of which are potentially very important to earthquake hazard research. One explanation of the heat flow paradox that has recently been suggested is that the actual mechanism of stick-slip observed on small rock samples in the laboratory (which originally led to the heat flow paradox, as a consequence of high stresses determined to be necessary for stick-slip) may not correspond to the mechanism of stick-slip in the earth, in part because of scaling problems, sample-machine interaction, and the existence of large variations in normal stress, and possibly even fault opening during the stick-slip process (Brune *et al.*, 1990, 1992; Anooshehpoor and Brune, 1994; Anooshehpoor and Brune, 1999). The normal vibrations mechanism was suggested by modeling of stick-slip between large blocks of foam rubber. In these models opening of the fault during stick-slip is clearly observed, resulting in a consequent reduction of frictional heat generation (Anooshehpoor and Brune, 1994).

Reduction in friction as a result of normal vibration has been investigated by many researchers over the past three decades (Tolstoi, 1967; Oden and Martins, 1985; Hess and Soom, 1991a, 1991b). Melosh (1996), based on the acoustic fluidization model, (Melosh, 1979), has also indicated that normal vibration may be a mechanism to reduce normal compressive stress and cause the dynamic weakening of the fault. If there is normal vibration on the fault, this may lead to anomalous P-wave radiation. Haskell (1964) suggested that tensile-like normal interface vibrations on the fault were required because the radiation of P-wave energy from large earthquakes was too high for pure shear faulting. Evidence from the ANZA array (Vernon *et al.*, 1989) and Guerrero strong motion array (Castro *et al.*, 1991) give some indication of anomalous P-wave excitation.

Normal compressive stress reduction with related normal motion along an interface between two different elastic media was first investigated by Weertman (1963; 1980). Based on the dynamic dislocation theory, a gliding edge dislocation along the interface could produce a change of tensile normal stress due to the material difference below and above the interface, while a change of shear stress was zero at a certain subsonic dislocation velocity. Usually, this subsonic dislocation velocity is limited between the *Rayleigh*-wave and shear-wave velocities of the softer material. The induced normal tensile stress adding to the normal compressive stress applied on the fault causes the reduction of normal compressive stress at the fault and sustains the slip dislocation propagation accompanied by a normal displacement motion. However, if the materials at the two sides of the interface are uniform, the normal stress induced by slip dislocation disappears immediately (Weertman, 1963; Aki and Richards, 1980). Physically, the coupling between the slip dislocation and change of normal stress is due to the asymmetry of material at the two sides of the fault (Andrews and Ben-Zion, 1997). Weertman (1963) pointed that this type of moving dislocation only exists in a narrow range of the material difference (a *19%* difference in the wave speeds of the two media). Andrews and Ben-Zion's (1997) calculations of a self-sustaining slip pulse on a fault between elastic media with wave speeds differing by *20%* confirms the prediction of Weertman (1980).

Based on the Weertman's dislocation approach, in this paper we use a 2-D lattice particle model to simulate the self-sustaining propagation of slip pulse involving interface separation on a planar fault between two elastic materials with different shear wave speeds. At first, by including climb dislocation (allowing opening of the fault) in Weertman's formulation, we find that self-sustaining dislocation pulses can propagate along the interface even if the difference in shear wave speeds in the two adjacent half-spaces exceeds *19%*. This result is important because in laboratory experiments we observe dynamic slip pulses that propagate along a planar fault between two large blocks of foam rubber with a shear wave speed difference of about *40%* (Anooshehpoor and Brune, 1999). We also find that by incorporating fault normal displacement discontinuity, the Weertman dislocation model degenerates into a dislocation model (Haskell, 1964) for identical half-spaces on both sides of the fault; and for a large difference in wave speeds in the two materials the slip pulse is somewhat similar to *Schallamach* detachment waves (Schallamach, 1971).

**Dislocation Theory:**

Dislocation theory from Weertman (1980) predicted that a steady state slip pulse can propagate along a dissimilar material interface governed by Coulomb friction. In his analysis, the fault normal motion was continuous across the material interface, that is, there is no interface separation was permitted. In this study, however, we demonstrate that, in the presence of interface separation, a self-sustaining slip pulse can propagate along an interface between two materials with arbitrarily different shear speeds.

Following Weertman (1980), the loading, particle motion and rupture propagation are in the x direction, and all variables are functions of *x*, *y*, and *t* only. Shear and dilatational wave speeds are $V_{si} = \sqrt{\mu_i / \rho_i}$ and $V_{pi} = \sqrt{(\lambda_i + 2\mu_i)/\rho_i}$ , where $\rho_i$ is mass density, $\lambda_i$, and $\mu_i$ are Lame's constants, and subscript *i* = 1, 2 denote the *i*[th] material. Shear and normal stresses on the fault plane are $\tau(x,t) = \sigma_{xy}(x, y = 0, t)$ and $\sigma(x,t) = \sigma_{yy}(x, y = 0, t)$. Applied shear stress and compressive normal stress at the remote boundaries are $\tau^\infty$ and $\sigma^\infty$. Fault parallel slip and fault normal dislocation (interface separation) are $U_x = u(x, y = 0^+, t) - u(x, y = 0^-, t)$ and $U_y = v(x, y = 0^+, t) - v(x, y = 0^-, t)$, respectively. If we define $B_x(X)$ and $B_y(X)$ as the shear and normal component distribution functions of an infinitesimal dislocation in the moving coordinate system of *X=x-ct*, here, *c* is the dislocation velocity, then $B_x(X)\delta X$ and $B_y(X)\delta X$ represent the shear and normal components of the dislocation in the interval between *X* and *X+δX*, respectively. For a uniform moving edge dislocation with velocity *c*, the relative particle velocities in the slip and normal directions are $\dot{U}_x = cB_x(X)$ and $\dot{U}_y = cB_y(X)$ , respectively. The shear and normal stress on the fault in the solution including normal dislocation are

$$\tau(X) = \tau^\infty + \frac{\overline{\mu}_1}{\pi} \int_{-\infty}^{+\infty} \frac{B_x(X^{'})}{X - X^{'}} dX^{'} - \mu^* B_y(X)$$

$$\sigma(X) = \sigma^\infty - \frac{\overline{\mu}_2}{\pi} \int_{-\infty}^{+\infty} \frac{B_y(X^{'})}{X - X^{'}} dX^{'} - \mu^* B_x(X)$$

(1)

where $\overline{\mu}_1$, $\overline{\mu}_2$ and $\mu^*$ are the composite elastic modules and are mismatch functions of the shear modulus $\mu_i$, density $\rho_i$, Poisson ratio $v_i$, and the dislocation velocity $c$ ( $i = 1,2$) (Weertman, 1980). The boundary conditions require that $\tau \le \tau^\infty$, which implies that there is no extra shear stress produced when fault interface separation, gives

$$\frac{\overline{\mu}_1}{\pi} \int_{-\infty}^{+\infty} \frac{B_x(X^{'})}{X - X^{'}} dX^{'} - \mu^* B_y(X) = 0$$

$$\sigma^\infty = \frac{\overline{\mu}_2}{\pi} \int_{-\infty}^{+\infty} \frac{B_y(X^{'})}{X - X^{'}} dX^{'} + \mu^* B_x(X)$$

(2)

**Figure 1** illustrates the normalized ratios $\overline{\mu}_1 / u_1$, $\overline{\mu}_2 / \mu_1$ and $\mu^* / \mu_1$ *versus* the normalized dislocation velocity $c / V_{s1}$ for the case of $\rho_1 = \rho_2$ and $V_{s1}/V_{s2} = 0.8$. It is seen that $\overline{\mu}_i$, and hence the long range shear stress ( $\sigma_{xy}^s = \overline{\mu}_1 / \pi \int_{-\infty}^{+\infty} B_x(X^{'}) / (X - X^{'}) dX^{'}$ ) and normal stress ( $\sigma_{yy}^n = \overline{\mu}_2 / \pi \int_{-\infty}^{+\infty} B_y(X^{'}) / (X - X^{'}) dX^{'}$ ) decrease as the dislocation velocity increases and reach to zero between $V_{R1}$, the slower *Rayleigh*-wave velocity and $V_{s1}$, the slower shear wave velocity. Also $\overline{\mu}_2$ decreases faster for the normal dislocation than that for the slip dislocation. The changes of the long-range shear and normal stresses arising from their corresponding dislocations are equal to zero at the zero points of $\overline{\mu}_1$ and $\overline{\mu}2$. The value of $\mu^*$ increases as $c$ increases and grows rapidly as the $c$ approaches $V_{s1}$. At certain rupture (dislocation) velocities in which $\overline{\mu}_1(c, \mu_i, \rho_i) = 0$ or $\overline{\mu}_2(c, \mu_i, \rho_i) = 0$, the values of $\mu^*$ are 3 to 4 times larger than that at $c = 0$. For an interface without material contrast, $\mu^*$ always equals zero for any given $c$, the rupture velocity. In general, a uniformly dislocation can not move in any arbitrary velocity of c under a given tectonic boundary with shear stress $\tau^\infty$ and normal compressive stress $\sigma^\infty$ around the fault, Otherwise, the dislocation motions are unstable, and there is no such tectonic boundary condition available to sustain such dislocations (both slip and normal components) to move uniformly and steadily. By setting $\bar{\mu}_1(c, \mu_i, \rho_i) = 0$ and $\bar{\mu}_2(c, \mu_i, \rho_i) = 0$ for slip and normal dislocations respectively, we get that the dislocation velocity of $c$ is dependent on the ratio of $V_{s1}/V_{s2}$ and is limited between the slower *Rayleigh*-wave and the slower shear-wave velocities. From equation (2), it is obvious that the condition required to sustain the interface separation for a given slip dislocation is $\bar{\mu}_1\bar{\mu}_2/\mu^* \geq 0$ (Adams, 1998). Previous discussions have shown that $\bar{\mu}_1 < 0$ and $\bar{\mu}_2 < 0$ are unacceptable conditions for dislocation motions under the tectonic boundary conditions.

## 2-D Lattice Particle Model:

In accordance with the objective of modeling rupture propagation along an interface between two elastic isotropic materials, we consider a two-dimensional triangular lattice particle model characterized by a pair potential. Particles interact with each other according to modified *Lennard-Jones* potential:

$$\phi(r) = \begin{cases} \frac{1}{2} k_i (r - r_0)^2 & r \leq 0 \\ \varepsilon_i \left[ \left( \frac{r_0}{r} \right)^{12} - 2 \left( \frac{r_0}{r} \right)^6 \right] & 0 \leq r < r_b \end{cases} \tag{3}$$

where $r_0$ is the rest length of equivalent *Hooke's* spring, $r_b$ is the cut-off distance equal to $1.112 r_0$, $k_i$ ($i=1,2$) are the linear spring constants related to the *Lame's* constants in which $\mu_i = \sqrt{3}/4 \, k_i$ with the *Poisson's* ratio of 0.25 (Hoover *et al.*, 1974), and $\varepsilon_i = r_0^2 k_i / 36$. Throughout the paper all results are expressed in terms of reduced units: the lattice length of $r_0$ is taken to be 1, the spring constant of $k_2$ equals 1 for a harder material and $0 < k_1 \leq 1$ for softer material, respectively. With the particle mass taken as the unit of mass, the triangular lattice particle has the density of $\rho = 2/2^{1/3}\sqrt{3}$, therefore, the resultant longitudinal, shear and the *Rayleigh* wave speeds for the harder material are

$$V_p = \sqrt{\frac{9}{8}} \approx 1.06, \quad V_s = \sqrt{\frac{3}{8}} \approx 0.61, \text{ and } V_R = 0.93 \times V_s \approx 0.57 \tag{4}$$

respectively.

With the assumption of material densities between the two elastic isotropic materials the same, $\bar{k}$, the interface elastic modulus (spring constant) of a mismatch function of $k_1$ and $k_2$, is taken as (Comninou, 1977b, Weertman, 1980)

$$\bar{k} = \frac{4k_1 k_2 \left[ k_2 \left( \kappa_2 + 1 \right) + k_1 \left( \kappa_1 + 1 \right) \right]}{\left[ k_2 \left( \kappa_1 + 1 \right) + k_1 \left( \kappa_2 + 1 \right) \right]^2 - \left[ k_2 \left( \kappa_1 - 1 \right) - k_1 \left( \kappa_2 - 1 \right) \right]^2} \tag{5}$$

where $\kappa_i = 3 - 4\nu_i, i = 1,2$, $\nu_i$ is the *Poisson* ratio. **Figure 2** shows that $\bar{k}$ varies as a function of ratio of $k_1 / k_2$ in which $k_1 \leq k_2 \leq 1$. It is clear that, when $0.5 < k_1 / k_2 \leq 1$, $\bar{k} < k_1 < k_2$. Therefore, the fault interface strength in this range is relative weaker than that of materials on the both sides of the fault. As we know that natural fault systems have interfaces that separate different materials. These are generated by damaged fault zone material or sometimes also by the existence of different rock bodies across the fault. Material interfaces are especially prominent in plate-bounding continental and subduction zone faults along which the largest earthquakes occur. The ranges of the typical material contrasts across the fault in the real earth depicted by ratio of shear speed differences are from 0.7 to 1.0. In our current model, because $V_i$, the shear speeds, are proportional to the square root of $k_i$, the spring constants, the ratios of $0.5 < k_1 / k_2 \leq 1$ approximately correspond to the ratios of shear speed differences of 0.7 to 1.0. Apparently, the fault interface properties, such as $\bar{k}$, the interface strength, and material contrasts described here by lattice particle approach are appropriated in describing real fault systems. Related to $\bar{k}$, the interface elastic modulus, the cohesive strength of the interface under the shear dominated deformation can be calculated as (Gao, *et al.*, 2001) $\tau_c + \sigma_c / \sqrt{3} = f_b / r_b$, where $\tau_c$ and $\sigma_c$ are the shear and normal stresses along the interface at cohesive limit. $r_b$ (~1.112$r_0$) is the break point in which $\phi''(r_b) = 0$ and $f_b = \phi'(r_b) = 0.0373 \bar{k}$, denoting the cohesive strength of a single bond. It is obvious that there is a strong coupling between shear and normal stresses during cohesive failure.

## Modeling:

The fault model is composed of two blocks with different elastic properties shown in **Figure 3**. The dimension of each block under study consists of *2000* lattice particles along the horizontal length defining the x-direction and 200 lattice particles along the vertical length defining the y-direction. In order to initiate rupture easily, a certain roughness at the leftmost edge of the fault (*200* particle distances) is added. Along the rest of the fault, the surfaces of the two sides of the fault are absolutely smooth. In our current study, the friction relation embedded on the fault is described by Coulomb friction law of $|\tau_s| = f \times \sigma_n$, where f, the friction coefficient gives in the range *f=0.6~0.85* (Byerlee, 1978), $\tau_s$ and $\sigma_n$ are the shear and normal stresses on the fault, respectively. A shear strain rate of *5\*10$^{-4}$* and a constant compressive strain of 0.002 are imposed on the outermost rows of particles in order to model the fault deformation driven by plate tectonic force. The top of the upper block is moving to the right and the bottom of the lower block is moving to the left.

For the numerical implementation of the model, a finite difference modified velocity *Verlet* algorithm (Allen and Tildesley, 1987) was been used, so that the new positions of the particles were calculated right after all the interactions had taken place.

## Numerical Results:

**1. Stick-slip motion and interface separation: Figure 4** presents the typical time histories of particle motions at a point on the fault with a material contrast of 0.3 ($V_{s1}/V_{s2}=0.7$). In general, the slip (frame 1) and normal (frame 2) displacements are similar to a ramp and pulse-shaped functions with a short rise-time, respectively. In comparison to the harder material (lower block), the particle motions both in the fault parallel and fault normal directions are much larger in the softer material (upper block). The interface separation is indicated by the difference of normal displacements at the two sides of the fault. The peak value of the normal displacement in the harder material is only *50%* of the peak value of the normal displacement in the softer material. If the material difference increases to *40%* or beyond, the normal particle motion in the harder material is so small that the particle motion is exactly the same as the *Schallmach* wave. The general features of these particle motions are also observed from foam rubber experiments (Anooshehpoor and Brune, 1999). From **Figure 4**, it also shows that the particle velocities and accelerations on the upper block (soft) are much larger than that in the lower block (hard). The fault normal components of particle velocities and accelerations are much larger than that of the fault parallel components too.

Additional calculation was carried out to explore the evolution of rupture propagation along the fault. The pulse-like particle motions grow sharper with a small increase in their amplitude as the rupture propagates away from the left edge of the rupture source as seen in **Figure 5**. This result is compatible with the results of Andrews and Ben-Zion (1997). A plausible explanation of such rupture evolution through the fault was discussed by Andrews and Ben-Zion (1997), through analysis of traveling waves along the fault, they found the rupture behaviors is influenced by travel waves radiated by rupture process. Close to the rupture initiation, the slower P-wave and the faster shear head wave affect the rupture motion, so that the P-wave prohibits the tensile separation of the fault. As the rupture propagates away from the initial source area, the different types of waves are separated from each other, and the rupture motion is controlled by the slower S-wave which promotes the tensile separation of the fault. In addition, our numerical study shows that pre-stored shear energy releases gradually to encourage the particle motion both in the tangential and normal directions as the rupture propagates away from the rupture initial point. The dynamic frictional stress usually drops to zero during the rupture, but the net shear stress required to initiate the rupture is much larger and remains on the fault plane. The net effect is to further accelerate the particle motion through the fault.

**Figure 6** illustrates a particular interesting instantaneous configuration of particle motion around the fault. Here, the dark circles indicate the particles in the softer material and the light circles indicate the particles in the harder material. The larger solid arrows indicate the relative fault motion. The smaller arrows indicate the opening starting point at the rupture front and healing point after the rupture, respectively. Obviously, during the rupture process, the particle motion configuration along the fault shows the particle displacement in the softer block having a larger displacement than that in the harder block. This instantaneous particle configuration also shows a wrinkle-like moving picture near the rupture front. This wrinkle-like surface pattern does not extend across the entire length of the fault but are rather localized. The particle arrays along the fault indicate a tensile motion involved in the rupture propagation. The calculated local rupture length of the particle pulse is about 20 ~ 25 lattice lengths.

**2. Rupture Mechanism:** Corresponding to the particle motion around the fault, the particle velocities are approximately pulse and sinusoidal functions in the slip and normal directions, respectively. These are consistent with the result of Andrews and Ben-Zion (1997) and the result predicted from dislocation theory (Weertman, 1980). **Figure 7** displays a particular instantaneous particle velocity field around the fault. The larger solid arrows indicate the relative motion direction of the fault, the lighter arrows indicate particle velocity field of the harder material, and darker arrows give particle velocity field of the softer material. The circle indicates the particle position, and the magnitude of each particle velocity is indicated by its vector length. From **Figure 7**, detailed analysis of the particle velocity field shows that, at the rupture front, the particle velocity is almost perpendicular to the fault, which indicates that the particles move upwards toward the softer medium. At the same time it is just at these points that the interface separation occurs.

Behind the rupture front, the opening process remains steady, the particle velocity is almost parallel to the fault, and the motion direction is asymmetric between the two blocks. In addition, the absolute value of the velocities between the two blocks is different; and the motion direction in softer medium is the same as the rupture direction. Also, the vectorized velocity field indicates that the maximum value of velocity is at the rupture front. When the interface is re-contacted, which corresponds to the healing phase, the interface particles move towards the fault plane, and, later on, towards the hard medium. The absolute value of velocity in the softer medium is larger. The slip motions start when interface separation occurs and stop when the interface re-contacts. From the particle velocity picture, it is clear that the tensile motion at the rupture front and the compressive motion after the rupture correspond to the interface separation and reconnecting, respectively. The stress distribution also gives the same picture.

**3. Near Fault Particle Motion:** In addition to the velocity field displayed in the **Figure 7**, **Figure 8** shows the synthetic seismograms at various distances away from the fault on the both sides of the fault. The motions on the both sides of the fault are predominately in the FN direction, and the synthetics show that the wavefields excited by the rupture source process consist of P, S, and Rayleigh waves, and the particle motions are dominated by Rayleigh waves with much longer duration of time. Strong surface wave energy emanates from the leading (rupture front) and trailing (healing point) edges of the slip zone and is directly related to the dynamic rupture process discussed above, as expected that the fault normal motions including interface separation play an important role in controlling rupture behavior. Corresponding to Figure 8, Figure 9 gives peak particle motion distributions across the fault. An asymmetric distribution pattern is so obvious that the particle motions in the fault normal and fault parallel directions in the soft material is much larger than that in the hard material. In addition, particle velocity of fault normal component is also much larger than that of fault parallel component. The results derived from Figure 9 also indicate that, as the distance away from the fault increases, the particle velocities both in the fault normal and fault parallel directions decrease rapidly. The physical mechanism has been discussed by Dunham and Archuleta (2004) based on the Fourier decomposition principle in which increasing the distance between the observer and the fault filters the high frequency components of the wavefield excited by the source process for sub-Rayleigh (sub-shear) ruptures. Moreover, as pointed out by Dunham and Archuleta (2004), a fundamental difficulty in source inversion we have to face is a resolvable problem when using records from subshear ruptures, even without the finite bandwidth limitation from the instrumental response and scattering along the ray path.

**4. Rupture velocity: Figure 10** shows the *x*-component particle velocity profile along the fault as a function of time. The strip pattern indicates the rupture propagates steadily along the entire fault. The slope of this time-distance curve gives the rupture velocity. *P*, *S* and *R* indicate the time-distance relations with the P-wave, S-wave and Rayleigh-wave velocities, respectively, of the slower medium. Measuring the slope of the particle velocity profile, we can see the rupture velocity lies between the Rayleigh-wave and S-wave velocities. The result is absolutely consistent with the result derived from dislocation theory (Weertman, 1980), in which the dislocation velocity is limited between the Rayleigh-wave and S-wave velocities of the slower medium when $V_{s1}/V_{s2}$ varies from *1.0* to a smaller value.

**5. Dynamic stresses: Figure 11** shows a typical shear stress variation along the fault before, during and after the rupture. From this figure we can see that, at the rupture front, there is a strong shear stress concentration before the rupture; the peak value could reach to 0.012. As the distance from the rupture front increases, the concentrated shear stress undergoes a $r^{-0.5}$ decay, rapidly tending to a static stress level. The dynamic stress drop is described by a shear stress decreasing from its critical value to zero temporarily, and locally, due to the opening behavior. The static stress drop is the difference of the shear stress before and after the rupture. In this figure we represent the dynamic stress drop and static stress drop as $\Delta\sigma_d$ and $\Delta\sigma_p$, respectively. Obviously, the variation of the shear stress through the rupture process shows a partial-stress-drop behavior, and the static stress drop is only about 20% to 25% of the dynamic stress drop. The particle motion along the fault undergoes a locking or stick → slip + opening → re-locking, or healing, during rupture. The partial stress drop feature of our model is a direct consequence of the fact that the particles re-connect after the rupture when they approach other particles at distances shorter than $r_0$. Although this is clearly a good representation of what occurs in the foam rubber model, it is, of course, an open question as to how closely this corresponds to the real earth.

An analysis of the stress distribution around the fault is an important issue in the evaluation of theories on dynamic rupture propagation. In this regard it should be recalled that the stress distribution around an opening crack in a discrete lattice differs from the linear elastic continuum description. Specifically, in a discrete lattice the crack tip stress is bounded; *i.e.,* there is no singularity. The maximum value that it can attain occurs at the rupture front and corresponds to the cohesive stress of the material. The stress concentration found at the rupture front exhibits a $r^{-0.5}$ stress dependency over a considerable range in $r$, in agreement with continuum prediction. Typical shear stress distribution observed for a dynamic rupture is displayed in **Figure 12** for a model with $4*10^5$ particles. The fault plane is indicated by solid line, and the upper block and lower block correspond to the soft and hard materials, respectively. As seen from **Figure 12**, the maximum shear stress is localized in a narrow region before the rupture front; and a shear stress concentration occurs just before the rupture front. At the rupture front, the shear stress pattern indicates a shear stress variation cross the fault with a small reduction in the softer medium. This phenomenon indicates the pre-rupture particle motion across the fault is very asymmetrical (England, 1965; Williams, 1959). In fact, dislocation theory predicted an asymmetrical shear distribution before the rupture front along the fault. Behind the rupture front, accompanied by fault opening, the shear stress has a big decrease.

**6. Ratio of $V_{s1}/V_{s2}$ and interface separation: Figure 13** shows the particle normal displacement pulse variations as $V_{s1}/V_{s2}$ increases from *0.5* to *0.95* at a constant normal compressive load. The normal pulse becomes sharper and narrower when $V_{s1}/V_{s2}$ increases from *0.5* to *0.95*. Their amplitude increases too. In addition, the normal displacements at the two sides of the fault get closer with each other as the ratio of $V_{s1}/V_{s2}$ becomes close to *1.0*. This implies that the interface separation behavior disappears gradually although the normal motion remains at the rupture front. The result is in agreement with the Haskell (1964) dislocation model in which the fault tensile motion is involved in the rupture propagation.

## Conclusions:

The study reported in this paper shows that a dynamic rupture along a dissimilar material interface is a rich phenomenon. The numerical results are, in general, consistent with the foam rubber experiments by Anooshehpoor and Brune (1999). Our main results may be summarized as follows:

1. With a material difference across the fault, the stick-slip motions accompanied by interface separation can exist under a shear dominated shear loading. The slip and normal displacements are ramp and pulse functions, respectively. The fault opening is indicated by the normal displacement difference between the two sides of the fault in which the particle motions are relative larger in the softer material. The discontinuities of the particle velocity both in the fault parallel and fault normal directions also indicate an asymmetrical radiation pattern about the fault.

2. The particles along the fault undergo a pulse-like motion, and the rupture pulse propagates with at the speed close to the *Rayleigh*-wave speed of the softer material, while the rupture direction is the same as the slip direction of the softer material. Fault opening is a localized process during the rupture process. The physical mechanism is somewhat similar to the *Schallamach* (1971) wave in which a soft rubber slides over a hard body. The major frictional energy is the work done in pulling the contact point apart from bonding surfaces as the rupture propagates.

3. The pulse-like particle motion derived from the inter-particle interaction (disconnecting and reconnecting) between the two sides of the fault results in short duration (rise time) of the rupture. In general, the localized rupture length is about *20 ~25* lattice length. The shear stress variation along the fault during the rupture reveals that the rupture motion is associated with a partial stress drop in which the transient dynamic stress drop is much higher than the final static stress drop.

4. The fault slip and opening exhibit a parametric dependence on the mismatch ratio of $V_{s1}/V_{s2}$. The ratio of $V_{s1}/V_{s2}$ determines the sharpness and the peak value of the fault normal particle motions at the two sides of the fault. As the $V_{s1}/V_{s2}$ approaches 1.0, the fault opening vanishes gradually, degenerating into a pure shear dislocation, in which normal displacements on both sides of the fault are the same and proportional to $log|X|$ at the rupture front $X \sim 0$ theoretically (Aki and Richards, 1980).

**Figure Captions:**

**Figure 1:** The composite elastic modules *versus* the dislocation velocity of *c*. The terms of $\bar{\mu}_1, \bar{\mu}_2$, and $\mu^*$ are given by Equation 10, 24 and 11 of Weertman (1980) for slip and normal dislocations, respectively.

**Figure 2:** $\bar{k}/k_2$ varies with $k_1/k_2$. $\bar{k}$, a composite interface spring constant, is derived from equation (5). If $0.5 < k_1/k_2 \leq 1$, then $\bar{k} < k_1 < k_2$.

**Figure 3:** Simulation geometry of 2D lattice particle model. The figure shows the blocks under a steady shear dominated loading. The spring constants in the upper and lower blocks are $k_1$ and $k_2$, respectively. The upper block is relatively soft compared to the lower block ($k_2 > k_1$).The interface strength, $\bar{k}$ is described by a composite elastic modulus of a mismatch function of $k_1$ and $k_2$.

**Figure 4:** Time histories of particle motions at a point on the fault. The upper and lower blocks correspond to the relative soft and hard materials, respectively. The interface separation is indicated by the difference of normal displacements at the two sides of the fault.

**Figure 5:** The evolution of the particle motions along the fault at the two sides of the fault. The results indicate, as the rupture propagates from the left to the right, the pulse-like particle motions become sharper and larger.

**Figure 6:** An instantaneous configuration of the particle motion between the soft (upper block) and the hard (lower block) materials. The larger solid arrows indicate a relative motion between the two sides of the fault. The small arrows indicate the direction of the particle motion at the rupture front (disconnecting point) and healing (reconnecting point) points. Near the rupture front, the particle motions indicate a local fault opening or interface separation.

**Figure 7:** An instantaneous particle velocity field around the fault during a rupture event. The large solid arrows indicate the relative motion of the fault, and the rupture propagates from the left to right. The small arrows indicate the directions of the particle motion. At the rupture front, the particle motions tend to move upwards toward the softer medium, and the interface separation occurs at the same time. Behind the rupture front, the particle velocity is almost parallel to the fault, and the interface separation remains steady. Corresponding to the interface reconnecting, the particles move toward the fault.

**Figure 8:** Synthetic seismograms of the particle velocity across the fault trace from the hard block to soft block. In the vertical coordinate, Going from negative to positive distance corresponds to going from the hard block, over the fault trace, to the soft block.

**Figure 9:** Peak particle velocities across the fault. Going from negative to positive distance corresponds to going from the hard block, over the fault trace, to the soft block. Solid curves denote the fault normal components, and dashed curves denote the fault parallel component.

**Figure 10:** The profile of the particle velocity along the fault *vs* time shows that the rupture propagates along the fault with a rupture velocity lies between Rayleigh-wave and shear-wave velocities of the softer material. The *R*, *S* and *P* with thick lines indicate the Rayleigh-wave, shear-wave and compressive-wave time-distance curves, respectively.

**Figure 11:** Shear stress distribution along the fault during a rupture. The open oval indicates a strong variation of shear stress during the interface reconnection (healing phase). The dashed line indicates an average static stress drop which is indicated by $\Delta\sigma_p$. The stress concentration is indicated by a large stress increase at the rupture front. The dynamic stress drop is given by $\Delta\sigma_d$.

**Figure 12:** Shear stress distribution around the fault during a rupture event. The fault is indicated by a solid line. The upper and lower blocks correspond to the relative low and high rigidities, respectively. (a) Instantaneous shear stress field associated with dynamic rupture propagation; and (b) contour-like shear stress distribution of (a).

**Figure 13:**    Mismatch ratio of shear wave velocities versus normal displacements at a point on the fault. The solid and dotted lines indicate the normal pulses in the softer and harder materials, respectively.

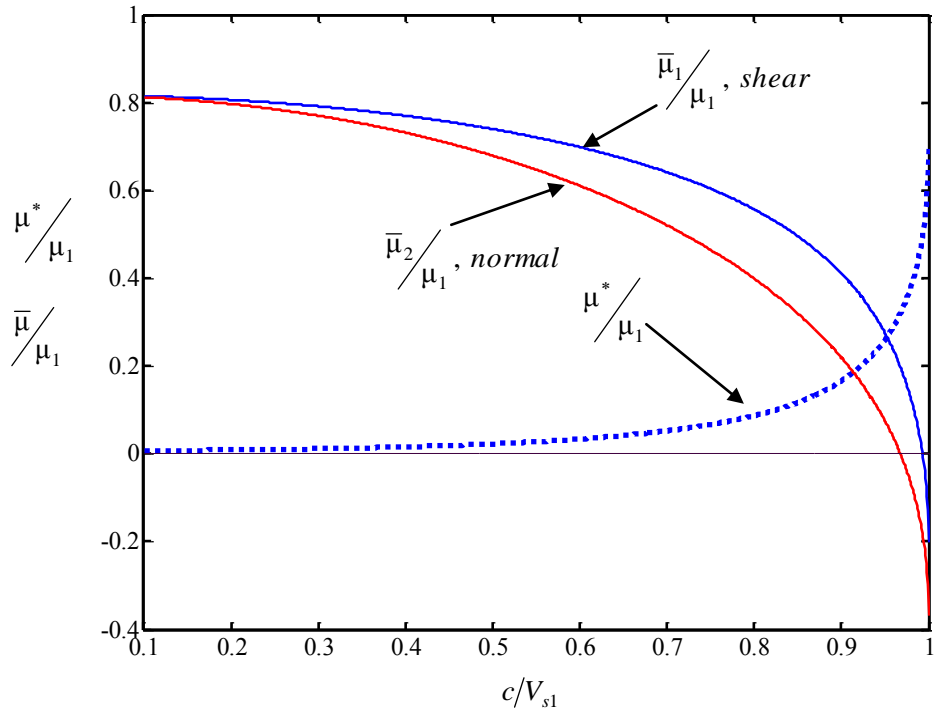

**Figure 1**

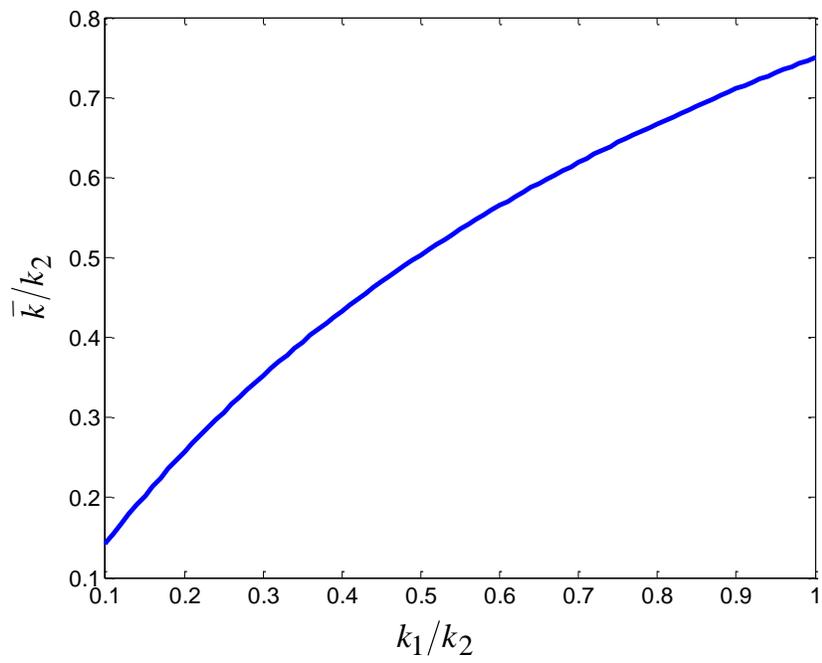

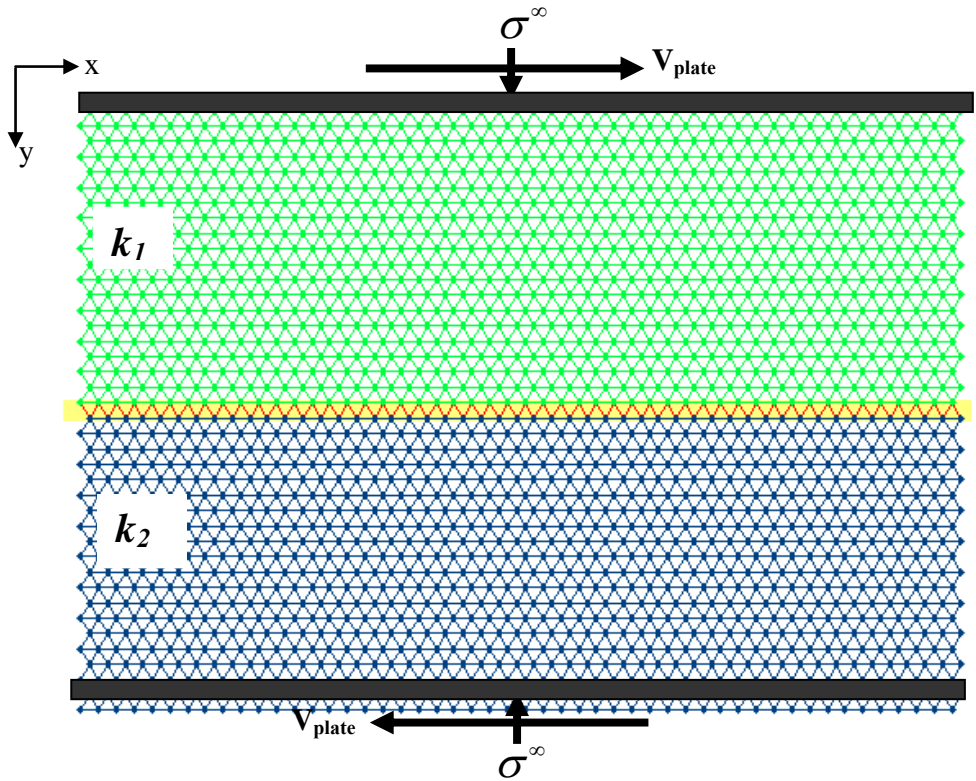

**Figure 3**

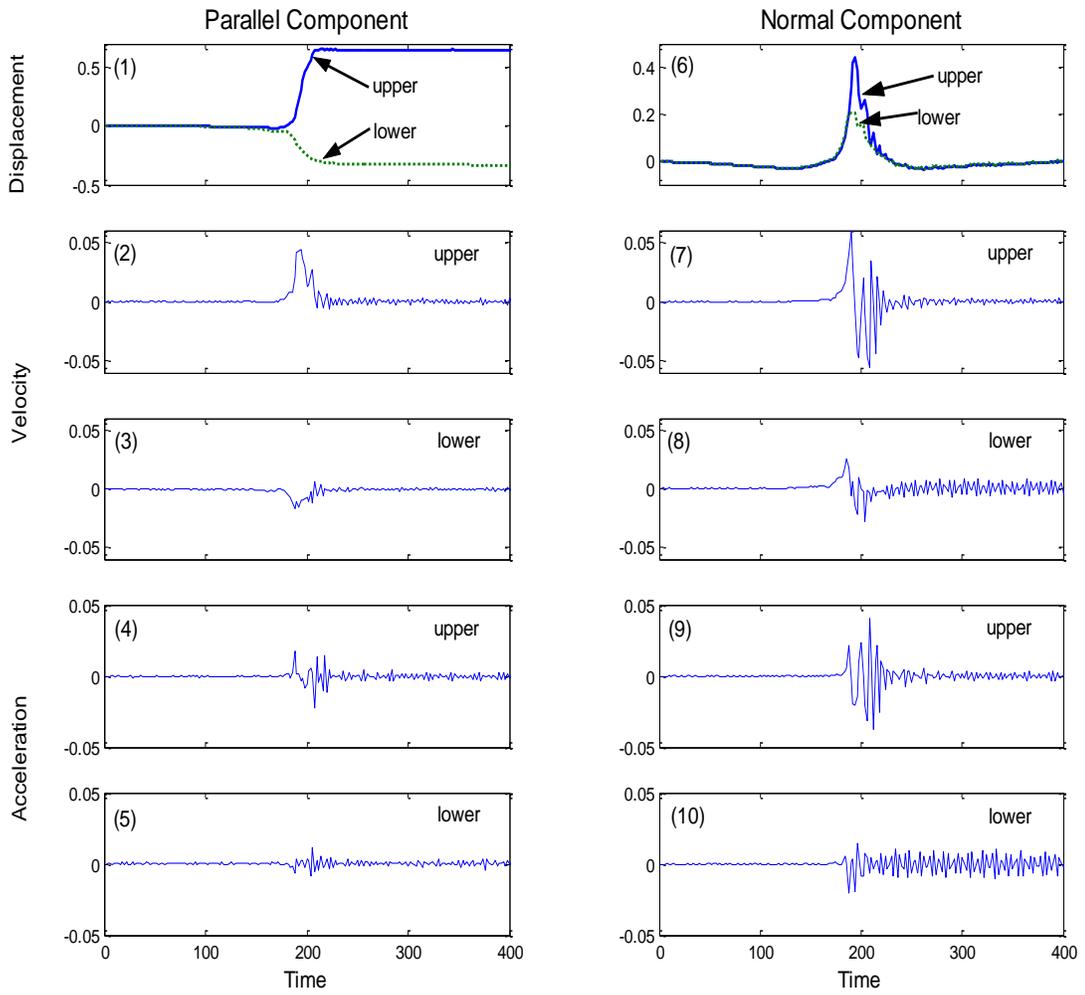

**Figure 4**

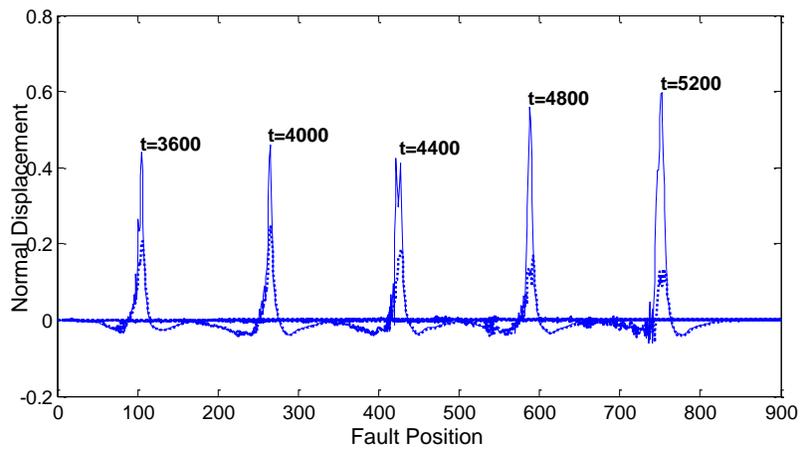

Figure 5

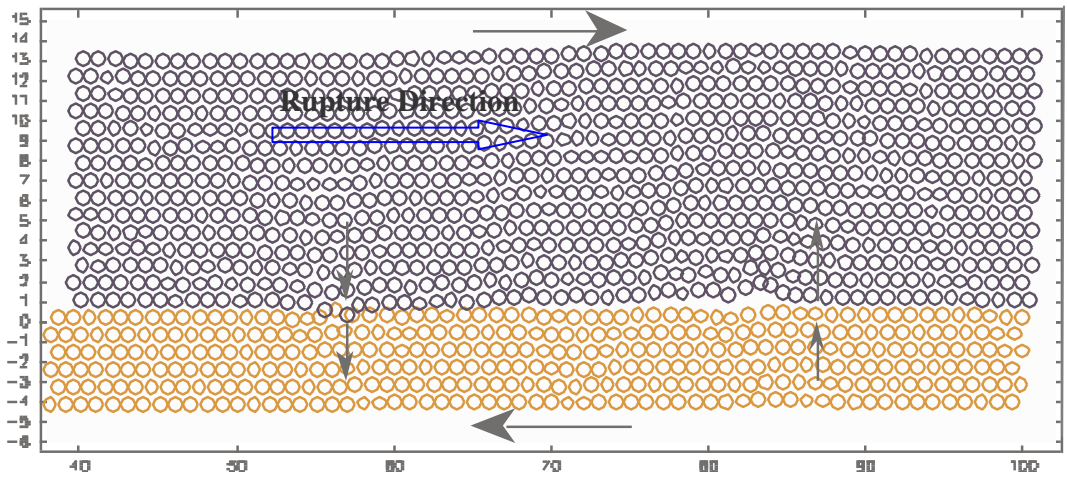

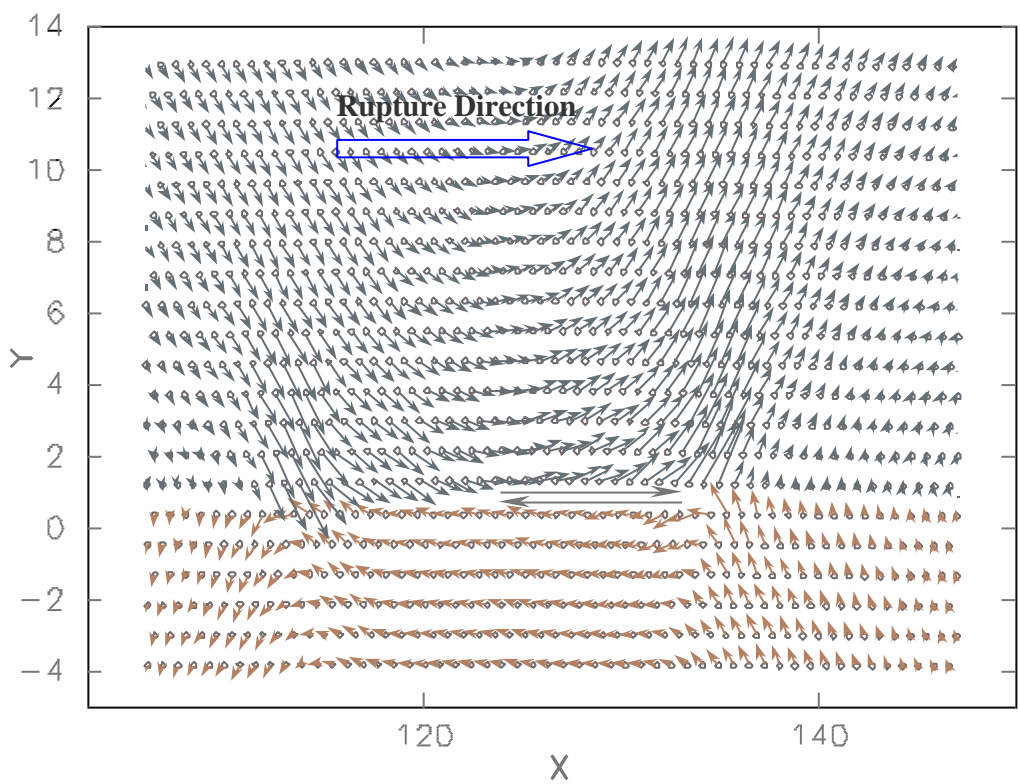

**Figure 7**

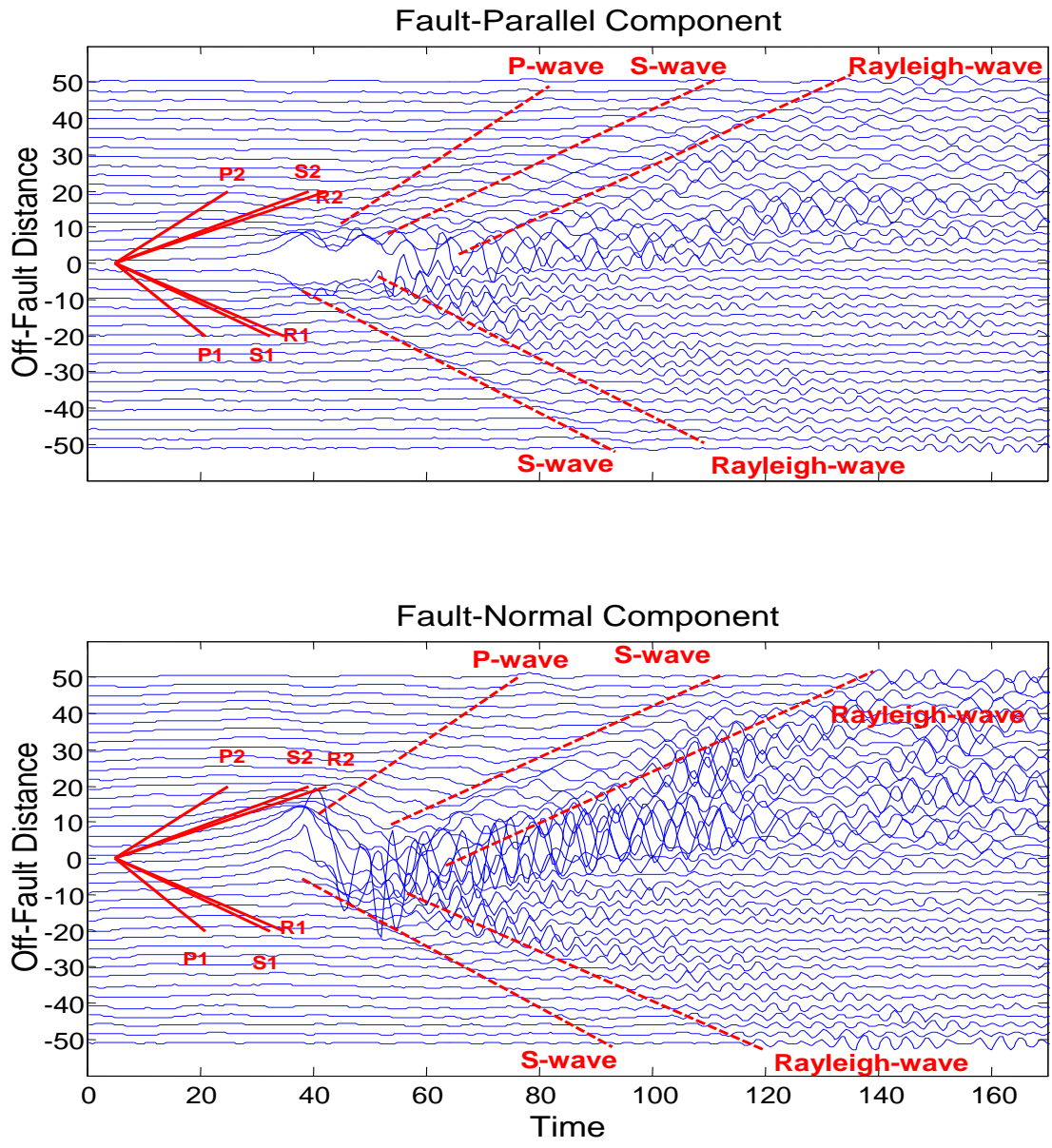

**Figure 8**

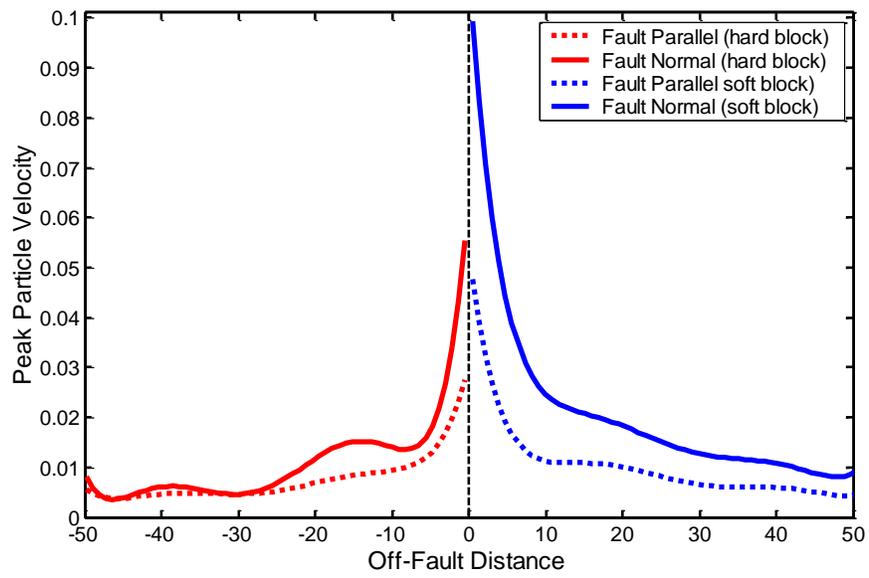

**Figure 9**

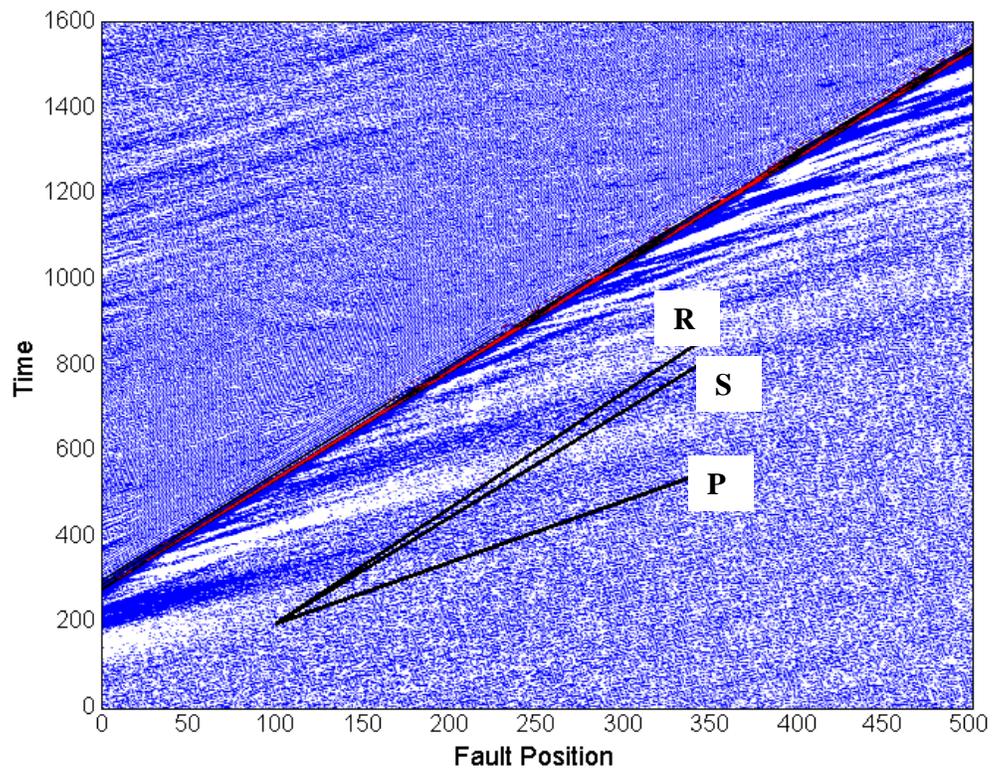

**Figure 10**

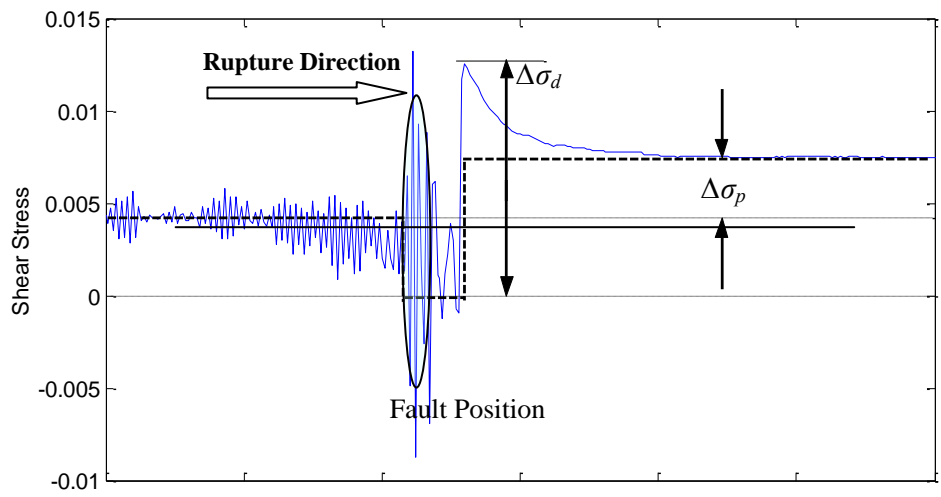

**Figure11**

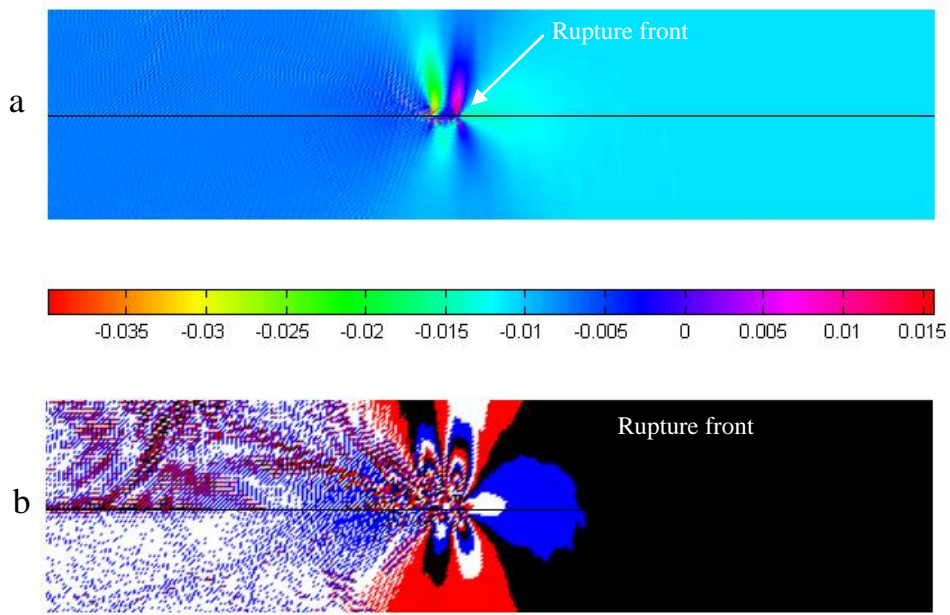

**Figure 12**

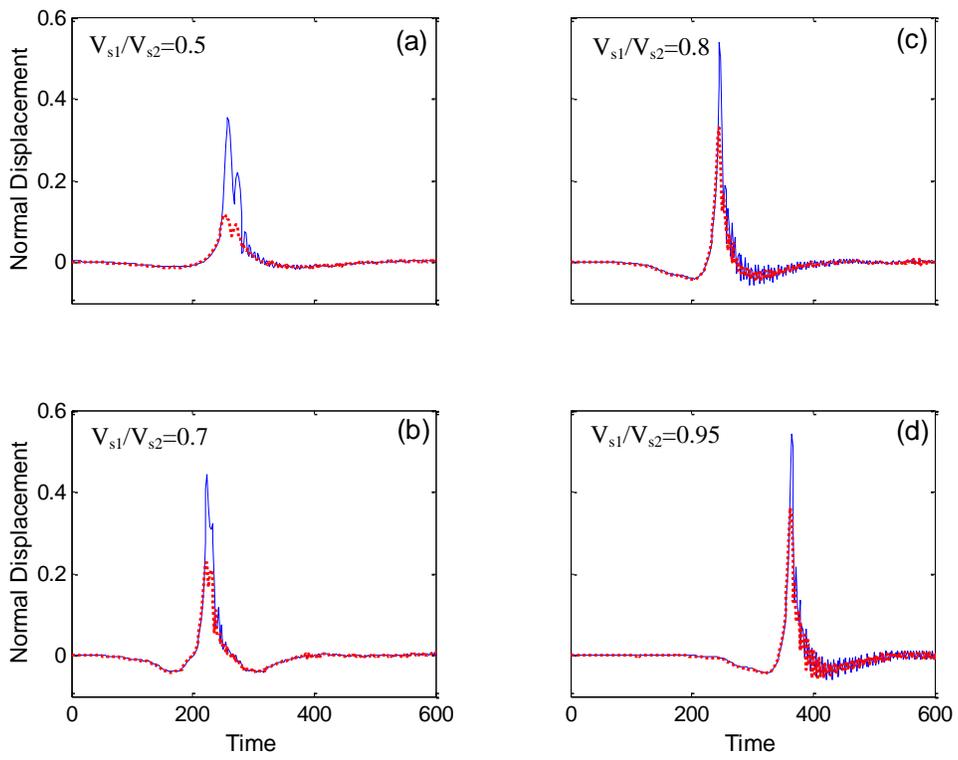

**Figure 13**